\documentclass[pdftex,twocolumn,epjc3]{svjour3}          

\RequirePackage{amsmath}
\RequirePackage{multirow}
\RequirePackage{subfigure}
\RequirePackage{hyperref}
\RequirePackage{color}

\allowdisplaybreaks[3]

\RequirePackage[T1]{fontenc}

\smartqed  

\RequirePackage{graphicx}

\journalname{Eur. Phys. J. C}

\begin{document}

\title{Predicted $\Xi_b(6087)^0$ and further predictions}

\author{Wei-Han Tan\thanksref{addr1} \and Hui-Min Yang\thanksref{addr2} \and Hua-Xing Chen\thanksref{addr1}
}

\institute{School of Physics, Southeast University, Nanjing 210094, China\label{addr1}
\and
School of Physics and Center of High Energy Physics, Peking University, Beijing 100871, China\label{addr2}}

\date{Received: date / Accepted: date}

\maketitle

\begin{abstract}
The methods of QCD sum rules and light-cone sum rules within the framework of heavy quark effective theory have been widely applied to study the singly heavy baryons, and especially, we have applied these methods to predict not only the mass and width of the $\Xi_b(6087)^0$ recently discovered by LHCb, but also its observation channel and its mass difference from the $\Xi_b(6100)^-$. We apply the same approach to perform a complete study on the $P$-wave bottom baryons of the $SU(3)$ flavor $\mathbf{\bar 3}_F$ and investigate their fine structure. Besides the $\Lambda_b(5912)^0$, $\Lambda_b(5920)^0$, $\Xi_b(6087)^0$, and $\Xi_b(6100)^-$, our results suggest the existence of the other two $\Lambda_b$ and two $\Xi_b$ baryons, whose widths are limited. They are the $\Lambda_b(J^P = 3/2^-)$ with the mass and width about $(5.93^{+0.13}_{-0.13}~{\rm GeV},\,0.0^{+12.0}_{-~0.0}~{\rm MeV})$, the $\Lambda_b(5/2^-)$ with $(5.94^{+0.13}_{-0.13}~{\rm GeV},\,\approx0~{\rm MeV})$, the $\Xi_b(3/2^-)$ with $\\(6.10^{+0.15}_{-0.10}~{\rm GeV},\, 1.4{^{+11.6}_{-~1.4}}~{\rm MeV})$, and the $\Xi_b(5/2^-)$ with $(6.11^{+0.15}_{-0.10}~{\rm GeV},\,1.0{^{+7.4}_{-1.0}}~{\rm MeV})$. Their mass splittings are calculated to be $M_{\Lambda_b(5/2^-)} - M_{\Lambda_b(3/2^-)} = 17\pm7~{\rm MeV}$ and $M_{\Xi_b(5/2^-)} - M_{\Xi_b(3/2^-)} = 14\pm7~{\rm MeV}$. All these baryons are explained as the $P$-wave bottom baryons of the $\rho$-mode, where the orbital excitation is between the two light quarks. However, the existence of this mode is still controversial, so their experimental searches can verify both our approach and the existence of the $\rho$-mode, which can significantly improve our understanding on the internal structure of hadrons.
\end{abstract}

$\\$
{\it Introduction} ---
The strong interaction holds quarks and gluons together inside the hadrons, which is similar in some aspects to the electromagnetic interaction that holds nucleons and electrons together inside the atoms. An ideal platform to study this is the singly bottom baryon system, which is composed of one bottom quark and two light quarks within the quark model picture. This baryon system, where the light quarks and gluons circle around the nearly static bottom quark, behaves as the QCD analogue of the hydrogen, where the electron circles around the nearly static proton~\cite{Korner:1994nh,Manohar:2000dt,Bianco:2003vb,Klempt:2009pi}. The electromagnetic interaction leads to the well-known fine structure of hydrogen spectra, and it is also interesting to investigate the fine structure of hadron spectra in the singly bottom baryon system caused by the strong interaction~\cite{pdg,Chen:2016spr,Copley:1979wj,Karliner:2008sv}.

In the past decade important experimental progresses were made in the field of singly bottom baryons, and a lot of singly bottom baryons were observed in particle experiments, such as the $\Lambda_b(5912)^0$ and $\Lambda_b(5920)^0$ discovered by the LHCb collaboration in 2012~\cite{LHCb:2012kxf}. These two states can be explained as the $P$-wave bottom baryons by Capstick and Isgur, since their predicted masses in 1986~\cite{Capstick:1985xss} are in a very good agreement with the experimental measurements~\cite{LHCb:2012kxf,CDF:2013pvu}. In 2021 the CMS collaboration further discovered the $\Xi_b(6100)^-$ in the $\Xi_b^-\pi^+\pi^-$ mass spectrum~\cite{CMS:2021rvl}, whose mass and width were measured to be:
\begin{eqnarray}
\nonumber \Xi_b(6100)^- &:& M = 6100.3 \pm 0.2 \pm 0.1 \pm 0.6{\rm~MeV} \, ,
\\      && \Gamma <1.9{\rm~MeV}~{\rm at}~95\%~{\rm CL} \, .
\end{eqnarray}
Later in 2022 we applied the QCD sum rule method to interpret the $\Lambda_b(5912)^0$, $\Lambda_b(5920)^0$, and $\Xi_b(6100)^-$ as the $P$-wave bottom baryons belonging to the $SU(3)$ flavor $\mathbf{\bar 3}_F$ representation, and moreover, we predicted and wrote in the abstract of Ref.~\cite{Yang:2022oog} that the $\Xi_b(6100)^-$ {\it ``has a partner state of $J^P=1/2^-$, labelled as $\Xi_b(1/2^-)$, whose mass and width are predicted to be $m_{\Xi_b(1/2^-)}=6.08^{+0.13}_{-0.11}$~GeV and $\Gamma_{\Xi_b(1/2^-)}=4^{+29}_{-~4}$~MeV, with the mass splitting $\Delta M=m_{\Xi_b(6100)}-m_{\Xi_b(1/2^-)}=9\pm3$~MeV. We propose to search for it in the $\Xi_c({1/2}^-)\to \Xi_b^{\prime}\pi$ decay channel.''} All these theoretical predictions are exactly the same as the experimental properties of the $\Xi_b(6087)^0$ discovered by LHCb in 2023~\cite{LHCb:2023zpu}, which was observed in the $\Xi_b^{\prime-} \pi^+$ mass distribution, with its mass and width measured to be:
\begin{eqnarray}
\nonumber  \Xi_b(6087)^0 &:& M = 6087.24 \pm 0.20 \pm 0.06 \pm 0.5{\rm~MeV} \, ,
\\     && \Gamma = 2.43 \pm 0.51 \pm 0.10{\rm~MeV} \, .
\end{eqnarray}

Besides the $\Lambda_b(5912)^0$, $\Lambda_b(5920)^0$, $\Xi_b(6087)^0$, and $\Xi_b(6100)^-$, there can exist more $P$-wave bottom baryons of the $SU(3)$ flavor $\mathbf{\bar 3}_F$. The successful predictions of our previous theoretical studies~\cite{Yang:2022oog} on the $\Xi_b(6087)^0$~\cite{LHCb:2023zpu} encourage us to further perform a complete study on these baryons. In this letter we shall work within the framework of heavy quark effective theory~\cite{Grinstein:1990mj,Eichten:1989zv,Falk:1990yz}, and apply the methods of QCD sum rules~\cite{Shifman:1978bx,Reinders:1984sr} and light-cone sum rules~\cite{Balitsky:1989ry,Braun:1988qv,Chernyak:1990ag,Ball:1998je} to calculate their mass spectra and decay properties, respectively. We refer to Refs.~\cite{Aliev:2018lcs,Azizi:2020azq,Yu:2021zvl} for more QCD sum rule studies. We shall predict two more $\Lambda_b$ baryons and two more $\Xi_b$ baryons, with limited widths and so capable of being observed in future particle experiments. Their masses and decay properties as well as their mass splittings within the same multiplets will be extracted for future experimental searches, as summarized in Table~\ref{tab:result}.

\begin{figure}[hbtp]
\begin{center}
\includegraphics[width=0.25\textwidth]{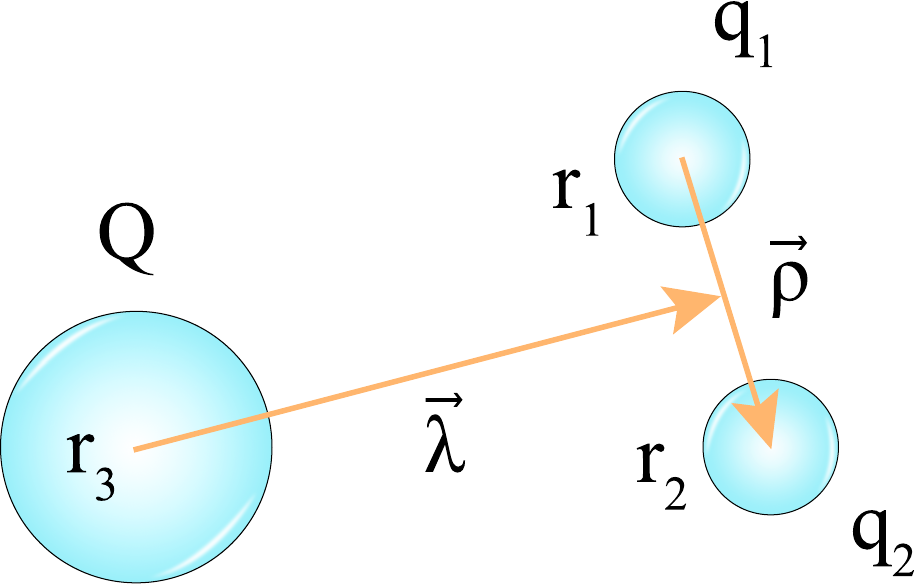}
\end{center}
\caption{Jacobi coordinates $\vec \lambda$ and $\vec \rho$ for the bottom baryon.}
\label{fig:Jacobi}
\end{figure}

In Ref.~\cite{Yang:2022oog} we explained the $\Lambda_b(5912)^0$, $\Lambda_b(5920)^0$, $\Xi_b(6087)^0$, and $\Xi_b(6100)^-$ as the $P$-wave bottom baryons of the $\rho$-mode, where the orbital excitation is between the two light quarks, as depicted in Fig.~\ref{fig:Jacobi}. This contradicts the expectation from some quark model calculations~\cite{Yoshida:2015tia,Nagahiro:2016nsx,Wang:2017kfr,Chen:2018orb,Kawakami:2019hpp,He:2021xrh}, so more experimental and theoretical studies are crucially demanded to further understand them. There is a relevant question on how to explain the five excited $\Omega_c$ baryons discovered by LHCb in 2017~\cite{LHCb:2017uwr}, given that at most four of them can be explained as the $P$-wave bottom baryons of the $\lambda$-mode~\cite{LHCb:2021ptx}. Two assignments are possible for the rest of them: either the orbital $1P$-wave excitation of the $\rho$-mode or the radial $2S$-wave excitation. Again, our QCD sum rule study of Ref.~\cite{Yang:2020zrh} supports the former assignment. Moreover, the two $\Lambda_b$ and two $\Xi_b$ baryons predicted in this letter also have the $\rho$-mode orbital excitation, so their experimental searches can verify not only our approach but also the existence of the $\rho$-mode. All their relevant theoretical and experimental studies can significantly improve our understanding on the internal structure of hadrons.

\begin{figure*}[hbtp]
\begin{center}
\includegraphics[width=0.8\textwidth]{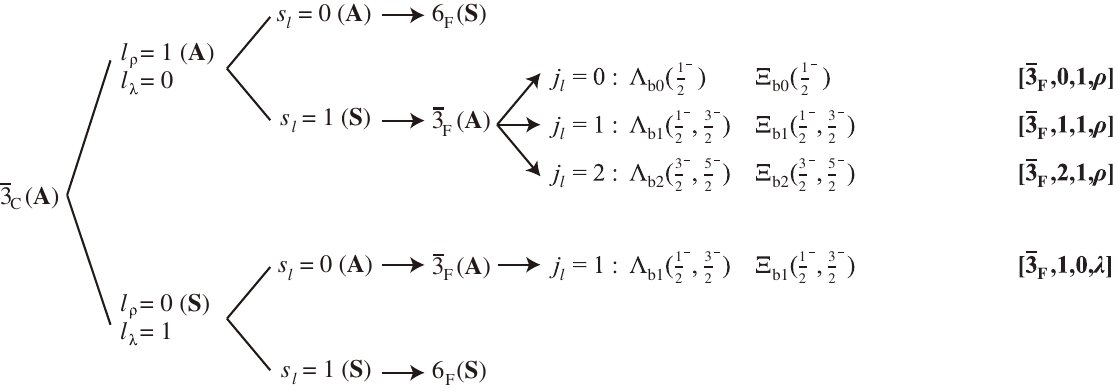}
\end{center}
\caption{Categorization of the $P$-wave singly bottom baryons belonging to the $SU(3)$ flavor $\mathbf{\bar 3}_F$ representation.}
\label{fig:pwave}
\end{figure*}

$\\$
{\it A global picture from the heavy quark effective theory} ---
As the first step, we classify the $P$-wave bottom baryons of the $SU(3)$ flavor $\mathbf{\bar 3}_F$ according to the heavy quark effective theory. A $P$-wave bottom baryon is composed of one $bottom$ quark and two light $up/down/strange$ quarks, and there can be as many as four internal symmetries between the two light quarks:
\begin{itemize}

\item The color structure of the two light quarks is always antisymmetric ($\mathbf{\bar 3}_C$).

\item The flavor structure of the two light quarks can be either antisymmetric ($SU(3)$ flavor $\mathbf{\bar 3}_F$) or symmetric ($SU(3)$ flavor $\mathbf{6}_F$).

\item The spin structure of the two light quarks can be either antisymmetric ($s_l \equiv s_{qq} = 0$) or symmetric ($s_l = 1$).

\item The orbital structure of the two light quarks can be either antisymmetric ($\rho$-mode with $l_\rho = 1$ and $l_\lambda = 0$) or symmetric ($\lambda$-mode with $l_\rho = 0$ and $l_\lambda = 1$), where $l_\rho$ denotes the orbital angular momentum between the two light quarks, and $l_\lambda$ denotes the orbital angular momentum between the bottom quark and the light quark system.

\end{itemize}
According to the Pauli principle, the total structure of the two light quarks should be antisymmetric. As shown in Fig.~\ref{fig:pwave}, we can categorize the $P$-wave bottom baryons into eight multiplets, four of which belong to the $SU(3)$ flavor $\mathbf{\bar 3}_F$ representation. The other four multiplets belong to the $SU(3)$ flavor $\mathbf{6}_F$ representation, and we have systematically studied them in Refs.~\cite{Yang:2020zrh,Yang:2021lce} within the framework of heavy quark effective theory. We denote them as $[flavor, j_l, s_l, \rho/\lambda]$, with $j_l = l_\lambda \otimes l_\rho \otimes s_l$ the total angular momentum of the light quark system. Each multiplet contains one or two bottom baryons, with the total angular momenta $J = j_l \otimes s_b = |j_l \pm 1/2|$.

$\\$
{\it Mass spectra from QCD sum rules} ---
In the heavy quark limit $m_b \to \infty$, the one-gluon-exchange interaction between the bottom quark and the light diquark is independent of the bottom quark mass, so the bottom quark looks like a static color source for the light diquark, and the hadron properties are mainly determined by the light quark system. In this framework the spin-$J$ baryon belonging to the multiplet $[flavor, j_l, s_l, \rho/\lambda]$ has the mass
\begin{equation}
m_J = m_b + \overline{\Lambda} + \delta m_J \, ,
\label{eq:mass}
\end{equation}
where $m_b$ is the bottom quark mass, $\overline{\Lambda}$ is the QCD sum rule result evaluated at the leading order, and
\begin{equation}
\delta m_J = -\frac{1}{2m_{b}}( K + d_{J}C_{mag}\Sigma ) \, ,
\label{eq:masscorrection}
\end{equation}
is the QCD sum rule result evaluated at the ${\mathcal O}(1/m_b)$ order, where $K$ and $\Sigma$ relate to the operators of the nonrelativistic kinetic energy and the chromomagnetic interaction, respectively. The coefficient $C_{mag} = [ \alpha_s(m_b) / \alpha_s(\mu) ]^{3/\beta_0}$ with $\beta_0 = 11 - 2 n_f /3$, and the other coefficient $d_{J}$ is defined as
\begin{equation}
d_{J=j_{l}-1/2} = 2j_{l}+2\, ,~~~~~
d_{J=j_{l}+1/2} = -2j_{l} \, .
\end{equation}
Hence, the $\Sigma$ term directly relates to the mass splitting within the same multiplet. This term is usually not large, so the fine structure of the singly bottom baryon system appears naturally through our approach.

We apply the QCD sum rule method within the framework of heavy quark effective theory to systematically calculate the mass spectra of the $P$-wave bottom baryons belonging to the $SU(3)$ flavor $\mathbf{\bar 3}_F$ representation. The obtained results are summarized in the third and fourth rows of Table~\ref{tab:result}. We clearly see from Eq.~(\ref{eq:mass}) that the baryon masses depend significantly on the bottom quark mass, so there is considerable uncertainty in our results for their absolute values. However, the mass splittings within the same doublet do not depend much on the bottom quark mass, which are produced quite well with much less uncertainty.

$\\$
{\it Decay properties from light-cone sum rules} ---
As shown in Table~\ref{tab:result}, we can not use only the mass spectra calculated through the QCD sum rule method to distinguish whether the $\Lambda_b(5912)^0$, $\Lambda_b(5920)^0$, $\Xi_b(6087)^0$, and $\Xi_b(6100)^-$ belong to the $[\mathbf{\bar 3}_F, 1, 1, \rho]$ doublet or the $[\mathbf{\bar 3}_F,1,0,\lambda]$ doublet. To better understand them we need to further investigate their decay properties, which can supply much more useful information. In this letter we apply the light-cone sum rule method within the framework of heavy quark effective theory to systematically study the decay properties of the $P$-wave bottom baryons belonging to the $SU(3)$ flavor $\mathbf{\bar 3}_F$ representation. The obtained results are summarized in the fifth and sixth rows of Table~\ref{tab:result}, where we have calculated their two/three-body $S$-wave decays and two-body $D$-wave decays into the ground-state bottom baryons together with light pseudoscalar mesons. Working at the renormalization scale 1~GeV, we have used the following QCD sum rule parameters for both the mass and width calculations~\cite{Ovchinnikov:1988gk,Colangelo:1998ga,Jamin:2002ev,Narison:2002woh,Ioffe:2002be,Gimenez:2005nt}:
\begin{eqnarray}
\nonumber \langle \bar qq \rangle &=& - (0.24 \pm 0.01)^3 \mbox{ GeV}^3 \, ,
\\ \nonumber \langle \bar ss \rangle &=& (0.8\pm 0.1)\times \langle\bar qq \rangle \, ,
\\ \langle g_s^2GG\rangle &=& (0.48\pm 0.14) \mbox{ GeV}^4\, ,
\label{eq:condensate1}
\\ \nonumber \langle g_s \bar q \sigma G q \rangle &=& M_0^2 \times \langle \bar qq \rangle\, ,
\\ \nonumber \langle g_s \bar s \sigma G s \rangle &=& M_0^2 \times \langle \bar ss \rangle\, ,
\\ \nonumber M_0^2 &=& 0.8 \mbox{ GeV}^2\, .
\end{eqnarray}
Besides, we have used the two PDG values~\cite{pdg}:
\begin{eqnarray}
&& m_s = (93^{+11}_{-~5}) \times 1.35 \mbox{ MeV} \, ,
\label{eq:condensate2}
\\ \nonumber && m_b = 4.18^{+0.03}_{-0.02} \mbox{ GeV} \, ,
\end{eqnarray}
for the running $strange$ and $bottom$ quark masses in the $\overline{\rm MS}$ scheme. Our results for the widths significantly depend on the inputting masses of $P$-wave bottom baryons. As shown in the third row of Table~\ref{tab:result}, our calculated masses have quite large uncertainties, so we use the experimental masses of $\Lambda_b(5912)^0$, $\Lambda_b(5920)^0$, $\Xi_b(6087)^0$, and $\Xi_b(6100)^-$ as inputs for the $[\mathbf{\bar 3}_F, 1, 1, \rho]$ doublet, while we still use our calculated masses as inputs for the other baryons.

As shown in Table~\ref{tab:result}, our results still can not distinguish whether the $\Lambda_b(5912)^0$ and $\Lambda_b(5920)^0$ belong to the $[\mathbf{\bar 3}_F, 1, 1, \rho]$ doublet or the $[\mathbf{\bar 3}_F,1,0,\lambda]$ doublet. However, the widths of $\Xi_b(6087)^0$ and $\Xi_b(6100)^-$ were respectively measured by LHCb~\cite{LHCb:2023zpu} to be $2.43 \pm 0.51 \pm 0.10$~MeV and $0.94 \pm 0.30 \pm 0.08$~MeV, so our results for the $[\mathbf{\bar 3}_F, 1, 0, \lambda]$ doublet are not easy to explain them, since the widths of $\Xi_b(1/2^-)$ and $\Xi_b(3/2^-)$ belonging to this doublet are respectively calculated to be $2800{^{+8600}_{-2800}}$~MeV and $330{^{+2010}_{-~330}}$~MeV. The large uncertainties of our theoretical results are mainly because that the relevant phase spaces are highly uncertain. If we use the experimental masses of $\Xi_b(6087)^0$ and $\Xi_b(6100)^-$ as inputs, the widths of $\Xi_b(1/2^-)$ and $\Xi_b(3/2^-)$ belonging to the $[\mathbf{\bar 3}_F, 1, 0, \lambda]$ doublet are respectively calculated to be $2000^{+2600}_{-1700}$~MeV and $86^{+110}_{-~86}$~MeV, so still it is not easy to use this doublet to explain the $\Xi_b(6087)^0$ and $\Xi_b(6100)^-$.

Our results suggest that the $[\mathbf{\bar 3}_F, 1, 1, \rho]$ doublet can be used to well explain the four excited bottom baryons $\Lambda_b(5912)^0$, $\Lambda_b(5920)^0$, $\Xi_b(6087)^0$, and $\Xi_b(6100)^-$ as a whole, and moreover, the $[\mathbf{\bar 3}_F, 2, 1, \rho]$ doublet can be used to predict two more $\Lambda_b$ baryons and two more $\Xi_b$ baryons. Paying attention to the experimental measurements of Refs.~\cite{LHCb:2012kxf,CDF:2013pvu,CMS:2021rvl,LHCb:2023zpu} (especially Fig.~2 of Ref.~\cite{CDF:2013pvu} and Fig.~2 of Ref.~\cite{LHCb:2023zpu}), we arrive at:
\begin{itemize}

\item $\Lambda_b^0(3/2^-)$. Its mass is about 5930~MeV and its width is nearly zero. It can be searched for in the $\Lambda_b\pi\pi$ decay channel.

\item $\Lambda_b^0(5/2^-)$. Its mass is about 5947~MeV and its width is also nearly zero.

\item $\Xi_b^0(3/2^-)$. Its mass is about 6102~MeV and its width is about 1.4~MeV. Its isospin partner $\Xi_b^-(3/2^-)$ has the mass about 6108~MeV. These two states can be searched for in the $\Xi_b\pi$ and $\Xi_b^*\pi$ decay channels.

\item $\Xi_b^0(5/2^-)$. Its mass is about 6116~MeV and its width is about 1.0~MeV. Its isospin partner $\Xi_b^-(5/2^-)$ has the mass about 6122~MeV. These two states can be searched for in the $\Xi_b\pi$ decay channel.

\end{itemize}

\begin{table*}[hbt]
\begin{center}
\renewcommand{\arraystretch}{1.4}
\caption{Mass spectra and decay properties of the $P$-wave bottom baryons belonging to the flavor $\mathbf{\bar 3}_F$ representation, calculated using the methods of QCD sum rules and light-cone sum rules within the framework of heavy quark effective theory. The partial decay widths are simply summed over to obtain the total decay widths. The uncertainties of some widths are quite large because the relevant phase spaces are highly uncertain. The $\Xi_b(6100)^-$ discovered by CMS~\cite{CMS:2021rvl} and the $\Xi_b(6095)^0$ discovered by LHCb~\cite{LHCb:2023zpu} are probably isospin partners.}
\begin{tabular}{ c | c | c | c | c | c | c}
\hline\hline
\multirow{2}{*}{Multiplet} & Baryon & Mass & Splitting & Partial Decay Width & Total Width  & \multirow{2}{*}{Candidate}
\\ & ($j^P$) & ({GeV}) & ({MeV}) & ({MeV}) & ({MeV}) &
\\
\hline\hline
\multirow{5}{*}{$[\mathbf{\overline{3}}_F, 1, 0, \lambda]$} & $\Lambda_b({1\over2}^-)$ & $5.91^{+0.17}_{-0.13}$& \multirow{2}{*}{$4\pm2$} &
$\begin{array}{c}
	\Gamma\left(\Lambda_b({1\over2}^-)\to \Sigma_b\pi\to \Lambda_b\pi\pi\right)=0\sim35000 \\
\end{array}$ &$0\sim35000$&--
\\ \cline{2-3}\cline{5-7}
& $\Lambda_b({3\over2}^-)$ & $5.91^{+0.17}_{-0.13}$ &&
$\begin{array}{c}
	\Gamma\left(\Lambda_b({3\over2}^-)\to \Sigma_b^{*}\pi\to \Lambda_b\pi\pi\right)=0\sim7200 \\
\end{array}$  &$0\sim7200$&--
\\ \cline{2-7}
&$\Xi_b({1\over2}^-)$ & $6.10^{+0.20}_{-0.10}$& \multirow{3}{*}{$4\pm2$} &
$\begin{array}{c}
	\Gamma\left(\Xi_b({1\over2}^-)\to \Xi_b^{\prime}\pi\right)=2800{^{+8600}_{-2800}} \\
\end{array}$ &$2800{^{+8600}_{-2800}}$&--
\\ \cline{2-3}\cline{5-7}
& \multirow{1}{*}{$\Xi_b({3\over2}^-)$} & $6.10^{+0.20}_{-0.10}$ &&
	$\begin{array}{c}
	\Gamma\left(\Xi_b({3\over2}^-)\to \Xi_b^{*}\pi\right)=330{^{+2010}_{-~330}} \\
	\Gamma\left(\Xi_b({3\over2}^-)\to \Xi_b^{\prime}\pi\right)=0.0{^{+1.9}_{-0.0}} \\
\end{array}$  &$330{^{+2020}_{-~330}}$&--
\\ \hline\hline
\multirow{2}{*}{$[\mathbf{\overline{3}}_F, 0, 1,\rho]$}&$\Lambda_b({1\over2}^-)$&$5.92^{+0.17}_{-0.19}$&\multirow{1}{*}{$-$}&
-- & -- &--
\\ \cline{2-7}
&$\Xi_b({1\over2}^-)$&$6.10^{+0.08}_{-0.08}$&\multirow{1}{*}{$-$}&
$\begin{array}{c}
	\Gamma\left(\Xi_b({1\over2}^-)\to\Xi_b\pi\right)=12000{^{+10000}_{-~9000}}\\
\end{array}$
&$12000{^{+10000}_{-~9000}}$&--
\\ \hline\hline
\multirow{5}{*}{$[\mathbf{\overline{3}}_F, 1, 1, \rho]$} & $\Lambda_b({1\over2}^-)$ & $5.92^{+0.13}_{-0.10}$& \multirow{2}{*}{$7\pm3$} &
$\begin{array}{c}
	\Gamma\left(\Lambda_b({1\over2}^-)\to \Sigma_b\pi\to \Lambda_b\pi\pi\right)={2{^{+5}_{-2}}} \times10^{-3} \\
\end{array}$ &${2{^{+5}_{-2}}} \times10^{-3}$&$\Lambda_b(5912)^0$
\\ \cline{2-3}\cline{5-7}
& $\Lambda_b({3\over2}^-)$ & $5.92^{+0.13}_{-0.10}$ &&
$\begin{array}{c}
	\Gamma\left(\Lambda_b({3\over2}^-)\to \Sigma_b^{*}\pi\to \Lambda_b\pi\pi\right)={5{^{+11}_{-~5}}} \times10^{-3} \\
\end{array}$  &${5{^{+11}_{-~5}}} \times10^{-3}$&$\Lambda_b(5920)^0$
\\ \cline{2-7}
&$\Xi_b({1\over2}^-)$ & $6.09^{+0.13}_{-0.12}$& \multirow{3}{*}{$7\pm3$} &
$\begin{array}{c}
	\Gamma\left(\Xi_b({1\over2}^-)\to \Xi_b^{\prime}\pi\right)=3.7{^{+9.5}_{-3.7}} \\
\end{array}$ &$3.7{^{+9.5}_{-3.7}}$&$\Xi_b(6087)^0$
\\ \cline{2-3}\cline{5-7}
& \multirow{1}{*}{$\Xi_b({3\over2}^-)$} & $6.10^{+0.13}_{-0.12}$ &&
$\begin{array}{c}
	\Gamma\left(\Xi_b({3\over2}^-)\to \Xi_b^{*}\pi\right)=0.64{^{+1.70}_{-0.64}} \\
	\Gamma\left(\Xi_b({3\over2}^-)\to \Xi_b^{\prime}\pi\right)={2{^{+4}_{-2}}} \times10^{-3} \\
\end{array}$  &$0.64{^{+1.70}_{-0.64}}$&
$\begin{array}{c}
\Xi_b(6095)^0\\
\Xi_b(6100)^-
\end{array}$
\\ \hline\hline
\multirow{9}{*}{$[\mathbf{\overline{3}}_F, 2, 1, \rho]$} & $\Lambda_b({3\over2}^-)$ & $5.93^{+0.13}_{-0.13}$& \multirow{2}{*}{$17\pm7$} &
$\begin{array}{c}
		\Gamma\left(\Lambda_b({3\over2}^-)\to \Sigma_b^{*}\pi\to \Lambda_b\pi\pi\right)=0.0^{+12.0}_{-~0.0}\\
\end{array}$ &$0.0^{+12.0}_{-~0.0}$&--
\\ \cline{2-3}\cline{5-7}
& $\Lambda_b({5\over2}^-)$ & $5.94^{+0.13}_{-0.13}$ &&
--  &$\approx0$&--
\\ \cline{2-7}
&$\Xi_b({3\over2}^-)$ & $6.10^{+0.15}_{-0.10}$& \multirow{5}{*}{$14\pm7$} &
	$\begin{array}{c}
	\Gamma\left(\Xi_b({3\over2}^-)\to \Xi_b\pi\right)=1.0{^{+8.3}_{-1.0}}\\
	\Gamma\left(\Xi_b({3\over2}^-)\to \Xi_b^{\prime}\pi\right)=0.00{^{+0.39}_{-0.00}}\\
	\Gamma\left(\Xi_b({3\over2}^-)\to \Xi_b^{*}\pi\right)=0.37{^{+3.26}_{-0.37}}\\
\end{array}$
 &$1.4{^{+11.6}_{-~1.4}}$&--
\\ \cline{2-3}\cline{5-7}
& \multirow{2}{*}{$\Xi_b({5\over2}^-)$} & $6.11^{+0.15}_{-0.10}$ &&
$\begin{array}{c}
	\Gamma\left(\Xi_b({5\over2}^-)\to \Xi_b\pi\right)=1.0{^{+7.2}_{-1.0}}\\
	\Gamma\left(\Xi_b({5\over2}^-)\to \Xi_b^{\prime}\pi\right)=0.00{^{+0.18}_{-0.00}}\\
	\Gamma\left(\Xi_b({5\over2}^-)\to \Xi_b^{*}\pi\right)=0.00{^{+0.01}_{-0.00}}\\
\end{array}$  &$1.0{^{+7.4}_{-1.0}}$&--
\\ \hline \hline
\end{tabular}
\label{tab:result}
\end{center}
\end{table*}

$\\$
{\it Conclusion} --- In summary, the successful predictions of our previous theoretical studies~\cite{Yang:2022oog} on the $\Xi_b(6087)^0$~\cite{LHCb:2023zpu} encourage us to perform a complete study on the $P$-wave bottom baryons of the $SU(3)$ flavor $\mathbf{\bar 3}_F$. According to the heavy quark effective theory, we categorize them into four multiplets: $[\mathbf{\bar 3}_F, 1, 0, \lambda]$, $[\mathbf{\bar 3}_F, 0, 1, \rho]$, $[\mathbf{\bar 3}_F, 1, 1, \rho]$, and $[\mathbf{\bar 3}_F, 2, 1, \rho]$. We apply the QCD sum rule method to calculate their mass spectra, and apply the light-cone sum rule method to calculate their decay properties. The obtained results are summarized in Table~\ref{tab:result}, where we have calculated their two/three-body $S$-wave decays and two-body $D$-wave decays into the ground-state bottom baryons together with light pseudoscalar mesons. More possible decay patterns will be investigated in our future study, where we shall provide additional details on our calculations.

Our results suggest that the $[\mathbf{\bar 3}_F, 1, 1, \rho]$ doublet can be used to explain the $\Lambda_b(5912)^0$, $\Lambda_b(5920)^0$, $\Xi_b(6087)^0$, and $\Xi_b(6100)^-$ as a whole. Besides, the $[\mathbf{\bar 3}_F, 2, 1, \rho]$ doublet can be used to further predict two more $\Lambda_b$ baryons and two more $\Xi_b$ baryons. We suggest the LHCb, CMS, and Belle-II collaborations to search for them in the $\Lambda_b\pi\pi$ and $\Xi_b^{(*)}\pi$ decay channels to see whether our approach can become one of the most effective ways to study the singly heavy baryon system. The above eight baryons all have the $\rho$-mode orbital excitation, whose existence is still controversial. Therefore, experimental searches of our predicted baryons can also verify the existence of the $\rho$-mode.

To end this letter, we note that the assignments proposed in this study are just possible explanations, and there exist many other possibilities. Further experimental and theoretical studies are crucially demanded to fully understand the singly heavy baryon system, whose beautiful fine structure is closely related to its rich internal structure. Recalling that the development of quantum physics is closely related to our understanding on the fine structure of hydrogen spectra, one naturally guesses that the currently undergoing studies on the singly heavy baryon system would not only improve our understanding on its internal structure, but also enrich our knowledge of the quantum physics.

\begin{acknowledgements}
This project is supported by
the National Natural Science Foundation of China under Grant No.~12075019,
the Jiangsu Provincial Double-Innovation Program under Grant No.~JSSCRC2021488,
and
the Fundamental Research Funds for the Central Universities.
\end{acknowledgements}

\bibliographystyle{elsarticle-num}
\bibliography{ref}

\end{document}